\documentclass[aps,twocolumn,showpacs, showkeys,prb]{revtex4}

\usepackage{graphicx}
\usepackage{graphicx,color,epsfig,rotate}
\usepackage{dcolumn}
\usepackage{bm}

\input{epsf.sty}

\begin{document}
\newcommand\hcone{$H_{c1}$}
\newcommand\hctwo{$H_{c2}$}
\newcommand\bacusio{BaCuSi$_2$O$_6$}

\title{Bose-Einstein Condensation of $S = 1$ Ni spin degrees of freedom in NiCl$_2$-4SC(NH$_2$)$_2$}
\author{V. S. Zapf,$^{1}$ D. Zocco,$^1$ M. Jaime,$^1$ N. Harrison,$^1$, A. Lacerda,$^1$, C. D. Batista,$^2$  A. Paduan-Filho,$^3$}

\affiliation{$^1$National High Magnetic Field Laboratory, Los
Alamos, NM \\ $^2$Condensed Matter and Statistical Physics, Los Alamos National Laboratory, Los Alamos, NM \\
$^3$ Instituto de Fisica, Universidade de Sao Paulo, Sao Paulo,
Brazil}

\date{\today}

\begin{abstract}

It has recently been suggested that the organic compound
NiCl$_2$-4SC(NH$_2$)$_2$ (DTN) exhibits Bose-Einstein Condensation
(BEC) of the Ni spin degrees of freedom for fields applied along the
tetragonal c-axis. The Ni spins exhibit 3D XY-type antiferromagnetic
order above a field-induced quantum critical point at $H_{c1} \sim
2$ T. The Ni spin fluid can be characterized as a system of
effective bosons with a hard-core repulsive interaction in which the
antiferromagnetic state corresponds to a Bose-Einstein condensate
(BEC) of the phase coherent $S = 1$ Ni spin system. We have
investigated the the high-field phase diagram and the occurrence of
BEC in DTN by means of specific heat and magnetocaloric effect
measurements to dilution refrigerator temperatures. Our results
indicate that a key prediction of BEC is satisfied; the magnetic
field-temperature quantum phase transition line $H_c(T)-H_{c1}
\propto T^\alpha$ approaches a power-law at low temperatures, with
an exponent $\alpha = 1.47 \pm 0.06$ at the quantum critical point,
consistent with the BEC theory prediction of $\alpha = 1.5$.

\end{abstract}


 \maketitle

In the past few years, a class of quantum spin systems that exhibit
a Bose-Einstein condensation in applied magnetic fields has been
receiving an increasing amount of attention. The compounds studied
previously consist of chains of Ni atoms,
\cite{Honda97,Tateiwa03,Tsujii04} planes of Cu dimers, (BaCuSiO$_6$)
\cite{Jaime04,Sebastian05} or 3-D coupled spin ladders (TlCuCl$_3$
and KCuCl$_3$). \cite{Ruegg03,Oosawa02} In these compounds, the spin
degrees of freedom are generally provided by either weakly
interacting $S=1$ Ni chains exhibiting a Haldane gap, or by a
lattice of dimerized $S = 1/2$ Cu$^{2+}$ ions. The spin singlet
ground state of these systems is separated from the lowest excited
state (triplet) by a finite energy gap at zero magnetic field. In
the presence of a magnetic field, the Zeeman term reduces the energy
of the $S = -1$ triplet state, until it reaches the energy of the
non-magnetic ground state at a field $H_{c1}$. Canted XY
antiferromagnetism is then observed between $H_{c1}$ and an upper
critical field $H_{c2}$. This XY antiferromagnetic phase of $S = 1$
spins has a U(1) symmetry and can thus be interpreted as a BEC. The
U(1) symmetry requires a spin environment that is axially symmetric
with respect to the applied field. Observation of a gapless
Goldstone mode (magnons) in the ordered phase could in principle
provide evidence of a broken continuous U(1) symmetry. Inelastic
neutron scattering measurements of the compound TlCuCl$_3$ were
proposed to be consistent with this interpretation. \cite{Ruegg03}
However, ESR measurements subsequently revealed significant magnetic
anisotropy in this compound inconsistent with a U(1) symmetry of the
spins. \cite{Glazkov04} In general, axial symmetry is not expected
for weakly coupled spin ladder and spin chain systems.

The compound NiCl$_2$-4SC(NH$_2$)$_2$ (DTN) \cite{PaduanFilho04} is
a new candidate for Bose-Einstein condensation of spins, and has
several features that make it unique among these BEC of spin
systems. In a similar manner to \bacusio{}, the tetragonal crystal
symmetry satisfies the U(1) symmetry requirement for BEC with the
additional feature that the symmetry can be tuned by rotating the
applied magnetic field. It has been predicted \cite{Batista05} that
an XY magnet should occur for fields along the c-axis, and an Ising
magnet for angles up to 40 degrees away from the c-axis. The spin
configuration is also different from previous BEC compounds,
consisting only of a spin-one state with no spin-singlet state. The
tetragonal spin lattice provided by the Ni$^{}$ ions is described by
the Hamiltonian:
\begin{equation}
{\cal H} = \sum_{{\bf j}, \nu} J_{\nu} {\bf S}_{\bf j} \cdot {\bf S}_{{\bf j} + {\bf e}_{\nu}}
+ D \sum_{\bf j} (S^z_{\bf j})^2,
\end{equation}
where $\nu=\{a,b,c \}$ and $J_a=J_b$ due to the tetragonal symmetry
of the crystal structure. The single ion anisotropy $D \sim 10$ K
\cite{PaduanFilho81,Batista05} splits the the Ni $S = 1$ spin state
into the $S_z = 0$ ground state and the $S_z = \pm 1$ excited
states.  Thus, for an applied magnetic field along the c-axis, the
level crossing occurs between two triplet states rather than a
triplet and a singlet as is the case for all other BEC systems
studied to date. Magnetization measurements \cite{PaduanFilho04}
have revealed AFM order between $H_{c1} \sim 2$ T and $H_{c2} \sim
12$ T with a maximum N\'{e}el temperature of 1.2 K. For a magnetic
field perpendicular to the $c$ axis, no transition is observed up to
$H = 15$ T since such a field mixes the $S_z = 0$ state with a
linear combination of the $S_z=\pm 1$ states, producing an effective
repulsion between energy levels that precludes any field-induced
quantum phase transition.

In this work, we investigate the temperature-magnetic field phase
diagram for $H~ ||~c$ via thermodynamic measurements. A key
prediction of the BEC theory is a power-law temperature dependence
of the phase transition line $H_c(T) - H_{c1} \propto T_c^\alpha$
where $\alpha = 1.5$. \cite{Affleck91} Previous studies on DTN and
other candidate BEC spin systems have found a wide variety of values
of $\alpha$, generally higher than the predicted value of 1.5. This
is due to the fact that the fits are performed at relatively high
temperatures, whereas the power-law universal behavior is only valid
for $T_c \rightarrow 0$. \cite{Sebastian05} Furthermore, the value
of $\alpha$ is very dependent on the value of $H_{c1}$ chosen for
the fit, which is difficult to extrapolate from higher temperature
data. A previous work found $\alpha = 2.6$ from fits to
magnetization data on DTN with most of the data lying between 0.5
and 1.5 K. \cite{PaduanFilho04} In this presetn study, we obtain
$\alpha$ from specific heat and magnetocaloric effect measurements
to dilution refrigerator temperatures. We determine the value of
$\alpha$ using the extrapolation method described by Sebastian
$et~al$.\cite{Sebastian05} The validity of this extrapolation has
been verified using Monte Carlo calculations of $H_c(T)$ and the
method has been successfully used to obtain the exponent for the
compound \bacusio{}.\cite{Sebastian05}

\epsfxsize=200pt
\begin{figure}[tbp]
\epsfbox{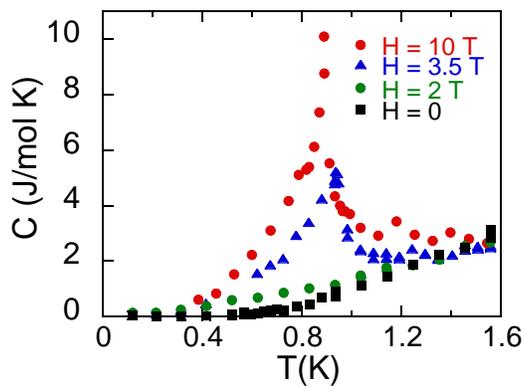} \caption{Specific heat $C$ versus
temperature $T$ of NiCl$_2$-4SC(NH$_2$)$_2$ for $H~||~c$ in applied
magnetic fields of 0, 2, 3.5, and 10 T. } \label{SpecificHeat}
\end{figure}

\epsfxsize=150pt
\begin{figure}[tbp]
\epsfbox{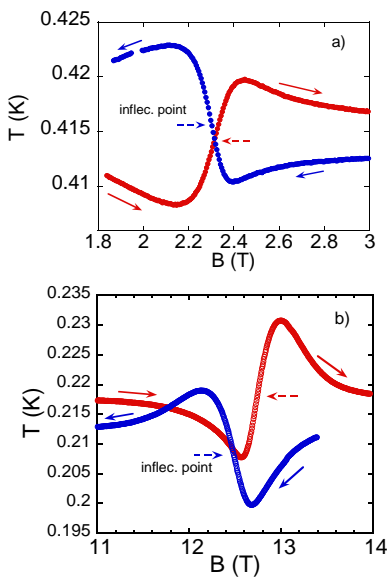} \caption{Magnetocaloric effect data
determined by monitoring $T$ while sweeping $B$ up and down. Solid
arrows indicate the direction of $B$ and dashed arrows indicate the
inflection point in $T$ versus $B$.}\label{MCE}
\end{figure}

Specific heat and the magnetocaloric effect (MCE) were measured in a
dilution refrigerator system with an 18 T magnet at the National
High Magnetic Field Laboratory in Los Alamos, NM. All measurements
were conducted on single crystals with the external field oriented
along the tetragonal c-axis. Specific heat (100 mK $\le T \le 1.5$
K) was determined by the quasi-adiabatic heat pulse relaxation
method, and the magnetocaloric effect was measured by sweeping the
field up an down at 0.1 T/min while monitoring the temperature
between 80 mK and 1.5 K.

\begin{figure}[tbp]
\centering \epsfbox{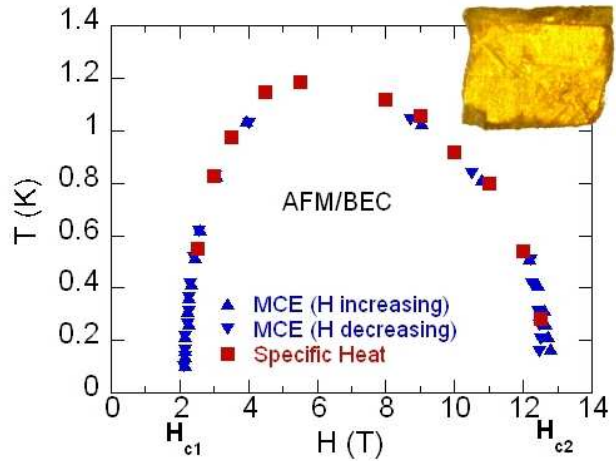} \caption{Temperature $T$ -
Magnetic field $H$ phase diagram from specific heat and
magnetocaloric effect data. Bose-Einstein Condensation (BEC)/canted
antiferromagnetic order (AFM) is thought to occur in the area under
the symbols. Top right: Photograph of the single crystal of DTN used in this work. The plane of the paper is
the $a-b$ plane of the crystal.} \label{PhaseDiagram}

\end{figure}

Distinct thermodynamic transitions can be observed in the specific
heat and magnetocaloric effect data at magnetic fields between 2.1
and 12.7 T. Representative data is shown in Figs \ref{SpecificHeat} and \ref{MCE}
and the phase diagram determined from these data is shown in Fig.
\ref{PhaseDiagram}. The specific heat data shown in Fig. \ref{SpecificHeat} exhibits
sharp peaks for $H = 3.5$ T and 10 T. These peaks
resemble the $\lambda$-like transition of the specific heat in
superfluid helium, an archetypal BEC system. Detailed theoretical
predictions for this system are necessary to make a precise
comparison. An equal entropy construction was used to determine the
midpoint of the transition. For transitions occurring at a given
temperature, the specific heat transition at high fields shows a
taller peak in the specific heat than the transition at low fields.

In the magnetocaloric effect data shown in Fig. \ref{MCE}, heating
is observed as the magnetic field is swept through the AFM
transition, surrounded by regions of cooling before and after the
transition. The region of cooling after the transition can be
attributed to a relaxation towards the bath temperature of the
system. However, the cooling preceding the transition must be due to
the sample. The inflection point of the $T(B)$ curve, corresponding
to the point of maximum heat, was identified as the phase
transition. The data at high field shown in Fig. \ref{MCE}(b) shows
hysteresis between the up and down sweep of the field. This
hysteresis is observed in the high field branch of the phase diagram
for temperatures less than 0.6 K, and is discussed in greater detail
later. The transition temperatures determined from specific heat and
MCE are in excellent agreement as shown in Fig. \ref{PhaseDiagram}.
A second order phase transition such as the AFM transition we are
expecting in this compound, should exhibit heating when entering the
ordered phase, and cooling when leaving it. However, all of the MCE
data shows heating in both directions when the field is swept up and
down. This may indicate a coupling to the lattice that results in an
apparent first-order phase transition.

\epsfxsize=240pt

The resulting phase diagram (Fig. \ref{PhaseDiagram}) shows an
ordered phase occurring between $H_{c1} \sim 2.1$ T and $H_{c2} \sim
12.7$ T with a maximum critical temperature of $T_c = 1.2$ K. The
values of $H_{c1}$ and $H_{c2}$ extracted from this phase diagram
are in agreement with those determined from magnetization to within
a few percent. \cite{PaduanFilho04} The discrepancy could be due to
slight orientation errors of the sample and the fact that in the
work by Paduan-Filho $et$ $al$, the onset of the transition is
determined from the peak in the first derivative of $M(H)$, whereas
a comparison with the midpoint of the transition in the specific
heat and MCE effect would require taking the inflection point, or
the peak in the second derivative of $M(H)$.

The values of $H_{c1}$ and $H_{c2}$ for $H~||~c$ can be obtained as
a function of $D$ and $J_{\nu}$ using a generalized spin-wave
approach \cite{Wang05,Batista05}:
\begin{equation}
g\mu_{B}H_{c1} = \sqrt{D^2-2zDJ}
\end{equation}
\begin{equation}
g\mu_{B}H_{c2} = D + 2zJ,
\end{equation}
where the coordination number $z = 6$. Here $J$ is the average of
$J_{\nu}$ over the three different directions $J =
\frac{1}{3}(2J_{a} + J_{c})$. The gyromagnetic ratio $g_c$ has been
determined from fits to the magnetization versus temperature as $g_c
= 2.26$ for $H || c$. Thus $D$ and $J$ can be solved for yielding
$D/k_B = 10.23$ K and $J/k_B = 0.77$ K. In contrast to \bacusio{},
the DTN compound is in a regime for which a two level description
does not work because $2zJ/D \sim 1$. In other words, the
fluctuations to the high energy $S_z=-1$ state play a role even at
very low energies. Consequently, the phase diagram is not symmetric
around the maximum critical field $H_{max}$ (see Fig.
\ref{PhaseDiagram}). The asymmetry is also manifested as a
difference between the size of the specific heat jumps at low and
high fields for a given critical temperature, as shown in Fig.
\ref{SpecificHeat}.

In addition to the asymmetry in the shape of the $H_c(T)$ curve, the
MCE data shows hysteresis for $T < 0.6$ K in the high field branch
of the phase diagram. Furthermore, an anomalous feature can be
observed in magnetization $M(H)$ data taken by Paduan-Filho $et$
$al$\cite{PaduanFilho04} at $T = 16$ mK and to a lesser degree at $T
= 0.6$ K. Between $H_{c1} \sim 2$ T and $H \sim 10$ T, the $M(H)$
data is linear as expected for a BEC. However, above $H = 10$ T, the
slope of $M$ versus $H$ increases until $M$ saturates above
$H_{c2}$. Thus, there is a region of anomalous behavior from
thermodynamic and magnetization measurements between $H = 10$ T and
$H_{c2}$, and below $T = 0.6$ K. One possible explanation is a
gradual expansion in the lattice beginning near $H = 10$ T. At high
magnetic fields approaching $H_{c2}$, the uniform magnetization
along the $c$-axis becomes large in competition with the
antiferromagnetic exchange interaction $J$. Thus, the system can
lower its energy by increasing the Ni-Ni distance to lower $J$. As a
result, $M(H)$ would saturate more rapidly with applied magnetic
field. The transition from the AFM phase back to the paramagnetic
phase at high fields would be accompanied by an abrupt cessation of
the lattice strain, consistent with the hysteretic first-order
transition observed in the MCE effect at high fields. Further
studies, such as magnetostriction are necessary to investigate this
possibility.

\epsfxsize=200pt
\begin{figure}[tbp]
\epsfbox{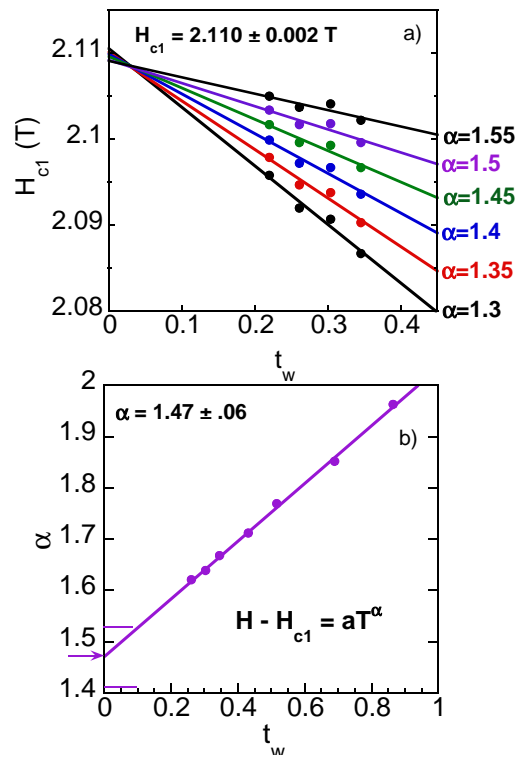} \caption{Estimation of the power-law
exponent $\alpha$ by fits over varying temperature ranges $t_w =
\Delta T/1.2$ K (see text). a) Fits of the expression $H_c(T) -
H_{c1} \propto T^\alpha$ were performed for various fixed values of
$\alpha$ at various temperature ranges, resulting in $H_{c1}$ a
function of temperature range $t_w$. The extrapolated value of
$H_{c1}$ at zero temperature is $H_{c1} = 2.110 \pm .002$ T. b)
Using the value of $H_{c1}$ determined in the left figure, fits to
$H_c(T) - H_{c1} \propto T^\alpha$ are performed for various
temperature range to determine $\alpha$, and $\alpha$ is shown
versus $t_w$. The extrapolated value of $\alpha$ at $t_w = 0$ is
indicated by the arrow, and the lines indicate the error bar. }
\label{Exponent}
\end{figure}

In order to study the occurrence of BEC of the Ni spins in DTN, we
examine the temperature dependence of the AFM phase transition line
near $H_{c1}$. As mentioned previously, a power-law temperature
dependence $H_c(T) - H_{c1} \propto T^\alpha$ where $\alpha = 1.5$
is predicted for a transition to a BEC phase. In our fits to the
$H_c(T)$ data for the compound DTN, we limit ourselves to
investigating the low field side of the phase diagram due to the
possibility of a magnetically induced strain phase occurring at high
fields. The value of $\alpha$ is closely dependent on the value of
$H_{c1}$, as well as on the temperature range of the fit. $\alpha$
is expected to approach 1.5 only as $T \rightarrow$ 0. The lack of
particle-hole symmetry does not allow us to use the the relation $t
\sim (1-h^2)^\nu g(h^2)$ ($\nu = 1/\alpha$, $t = T_c/T_{max}$ and $h
= \frac{H_{max}-H}{H_{max}-H_{c1}}$) that was exploited to determine
the exponent $\alpha$ for \bacusio{}.\cite{Sebastian05} Therefore,
we use the simplest power law expression: $H_c(T) -H_{c1} \propto
T^{\alpha}$.

As a first step, we fix $\alpha$ and fit the $H_{c}(T)$ data to
determine $H_{c1}$. This fit is performed for data up to a maximum
temperature where the window size $t_w = T_{max}/1.2 K$. The fits
are then repeated for different trial values of $\alpha$. The
results of these fits are shown in Fig. \ref{Exponent} (left). For
each value of $\alpha$, the $H_{c1}$ values as a function of the
window size $t_w$ can be fit to a straight line and extrapolated to
a window size $t_w = 0$. The extrapolations of $H_{c1}(t_w)$ for the
different values of $\alpha$ converge near $t_w = 0$, resulting in a
mean value of $H_{c1} = 2.110 \pm 0.002$ T.

In the expression $H_c(T) - H_{c1} \propto T^\alpha$, $H_{c1}$ can
now be fixed at $2.11$ T and the $H_c(T)$ data is fit to determine
$\alpha$ as a function of window size as shown in Fig.
\ref{Exponent} (right). For $t_w = 0$, $\alpha$ extrapolates to
$\alpha = 1.47 \pm 0.06$ where the error bar in $\alpha$ results
from the error bar in $H_{c1}$. This value is within experimental
error of the predicted exponent $\alpha = 1.5$ for a Bose-Einstein
condensate. Thus our data is consistent with a Bose-Einstein
condensation of the Ni spin degrees of freedom.

In summary, we have mapped the high-field phase diagram of DTN using
specific heat and magnetocaloric effect data, yielding a magnetic
ordered state between $H_{c1} = 2.1$ T and $H_{c2} = 12.7$ T with a
maximum critical temperature of 1.2 K. An anomalous region of the
phase diagram is observed between $H \sim 10$ and $12.7$ T and below
$T \sim 0.6$ K possibly associated with lattice strain. The phase
transition line near the quantum critical point at $H_{c1}$ can be
fit by the expression $H_c(T) - H_{c1} \propto T^\alpha$, resulting
in a value of $\alpha = 1.47 \pm 0.06$, which is consistent with the
predicted exponent equal to $1.5$ for a quantum phase transition to
a BEC.

\begin{acknowledgments}
This work was supported by the DOE and the NSF through the National
High Magnetic Field Laboratory and the LANL Director-Funded
Postdoctoral program. A.P.F. acknowledges support from CNPq
(Conselho Nacional de Desenvolvimento Científico e Tecnológico,
Brazil).

\end{acknowledgments}


\begin{thebibliography}{13}
\expandafter\ifx\csname
natexlab\endcsname\relax\def\natexlab#1{#1}\fi
\expandafter\ifx\csname bibnamefont\endcsname\relax
  \def\bibnamefont#1{#1}\fi
\expandafter\ifx\csname bibfnamefont\endcsname\relax
  \def\bibfnamefont#1{#1}\fi
\expandafter\ifx\csname citenamefont\endcsname\relax
  \def\citenamefont#1{#1}\fi
\expandafter\ifx\csname url\endcsname\relax
  \def\url#1{\texttt{#1}}\fi
\expandafter\ifx\csname urlprefix\endcsname\relax\def\urlprefix{URL
}\fi \providecommand{\bibinfo}[2]{#2}
\providecommand{\eprint}[2][]{\url{#2}}

\bibitem[{\citenamefont{Honda et~al.}(1997)\citenamefont{Honda, Katsumata,
  Katori, Yamada, Ohishi, Manabe, and Yamashita}}]{Honda97}
\bibinfo{author}{\bibfnamefont{Z.}~\bibnamefont{Honda}},
  \bibinfo{author}{\bibfnamefont{K.}~\bibnamefont{Katsumata}},
  \bibinfo{author}{\bibfnamefont{H.~A.} \bibnamefont{Katori}},
  \bibinfo{author}{\bibfnamefont{K.}~\bibnamefont{Yamada}},
  \bibinfo{author}{\bibfnamefont{T.}~\bibnamefont{Ohishi}},
  \bibinfo{author}{\bibfnamefont{T.}~\bibnamefont{Manabe}}, \bibnamefont{and}
  \bibinfo{author}{\bibfnamefont{M.}~\bibnamefont{Yamashita}},
  \bibinfo{journal}{J. Phys.: Condens. Matter} \textbf{\bibinfo{volume}{9}},
  \bibinfo{pages}{L83} (\bibinfo{year}{1997}).

\bibitem[{\citenamefont{Tateiwa et~al.}(2003)\citenamefont{Tateiwa, Hagiwara,
  Aruga-Katori, and Kobayashi}}]{Tateiwa03}
\bibinfo{author}{\bibfnamefont{N.}~\bibnamefont{Tateiwa}},
  \bibinfo{author}{\bibfnamefont{M.}~\bibnamefont{Hagiwara}},
  \bibinfo{author}{\bibfnamefont{H.}~\bibnamefont{Aruga-Katori}},
  \bibnamefont{and} \bibinfo{author}{\bibfnamefont{T.~C.}
  \bibnamefont{Kobayashi}}, \bibinfo{journal}{Physica B}
  \textbf{\bibinfo{volume}{329-333}}, \bibinfo{pages}{1209}
  (\bibinfo{year}{2003}).

\bibitem[{\citenamefont{Tsujii et~al.}()\citenamefont{Tsujii, Honda, Andraka,
  Katsumata, and Takano}}]{Tsujii04}
\bibinfo{author}{\bibfnamefont{H.}~\bibnamefont{Tsujii}},
  \bibinfo{author}{\bibfnamefont{Z.}~\bibnamefont{Honda}},
  \bibinfo{author}{\bibfnamefont{B.}~\bibnamefont{Andraka}},
  \bibinfo{author}{\bibfnamefont{K.}~\bibnamefont{Katsumata}},
  \bibnamefont{and} \bibinfo{author}{\bibfnamefont{Y.}~\bibnamefont{Takano}},
  \bibinfo{note}{cond-mat/0409190]}.

\bibitem[{\citenamefont{Jaime et~al.}(2004)\citenamefont{Jaime, Correa,
  Harrison, Batista, Kawashima, Kazuma, Jorge, Stern, Heinmaa, Zyvagin
  et~al.}}]{Jaime04}
\bibinfo{author}{\bibfnamefont{M.}~\bibnamefont{Jaime}},
  \bibinfo{author}{\bibfnamefont{V.~F.} \bibnamefont{Correa}},
  \bibinfo{author}{\bibfnamefont{N.}~\bibnamefont{Harrison}},
  \bibinfo{author}{\bibfnamefont{C.~D.} \bibnamefont{Batista}},
  \bibinfo{author}{\bibfnamefont{N.}~\bibnamefont{Kawashima}},
  \bibinfo{author}{\bibfnamefont{Y.}~\bibnamefont{Kazuma}},
  \bibinfo{author}{\bibfnamefont{G.~A.} \bibnamefont{Jorge}},
  \bibinfo{author}{\bibfnamefont{R.}~\bibnamefont{Stern}},
  \bibinfo{author}{\bibfnamefont{I.}~\bibnamefont{Heinmaa}},
  \bibinfo{author}{\bibfnamefont{S.~A.} \bibnamefont{Zyvagin}},
  \bibnamefont{et~al.}, \bibinfo{journal}{Phys. Rev. Lett}
  \textbf{\bibinfo{volume}{93}}, \bibinfo{pages}{087203}
  (\bibinfo{year}{2004}).

\bibitem[{\citenamefont{Sebastian et~al.}()\citenamefont{Sebastian, Sharma,
  Jaime, Harrison, Correa, Balicas, Kawashima, Batista, and
  Fisher}}]{Sebastian05}
\bibinfo{author}{\bibfnamefont{S.~E.} \bibnamefont{Sebastian}},
  \bibinfo{author}{\bibfnamefont{P.~A.} \bibnamefont{Sharma}},
  \bibinfo{author}{\bibfnamefont{M.}~\bibnamefont{Jaime}},
  \bibinfo{author}{\bibfnamefont{N.}~\bibnamefont{Harrison}},
  \bibinfo{author}{\bibfnamefont{V.}~\bibnamefont{Correa}},
  \bibinfo{author}{\bibfnamefont{L.}~\bibnamefont{Balicas}},
  \bibinfo{author}{\bibfnamefont{N.}~\bibnamefont{Kawashima}},
  \bibinfo{author}{\bibfnamefont{C.~D.} \bibnamefont{Batista}},
  \bibnamefont{and} \bibinfo{author}{\bibfnamefont{I.~R.}
  \bibnamefont{Fisher}}, \bibinfo{note}{submitted to Phys. Rev. Lett;
  [cond-mat/0408100]}.

\bibitem[{\citenamefont{R\"uegg et~al.}(2003)\citenamefont{R\"uegg, Cavadini,
  Furrer, G\"udel, Kr\"amer, Mutka, Wildes, Habicht, and
  Vorderwisch}}]{Ruegg03}
\bibinfo{author}{\bibfnamefont{C.}~\bibnamefont{R\"uegg}},
  \bibinfo{author}{\bibfnamefont{N.}~\bibnamefont{Cavadini}},
  \bibinfo{author}{\bibfnamefont{A.}~\bibnamefont{Furrer}},
  \bibinfo{author}{\bibfnamefont{H.-U.} \bibnamefont{G\"udel}},
  \bibinfo{author}{\bibfnamefont{K.}~\bibnamefont{Kr\"amer}},
  \bibinfo{author}{\bibfnamefont{H.}~\bibnamefont{Mutka}},
  \bibinfo{author}{\bibfnamefont{A.}~\bibnamefont{Wildes}},
  \bibinfo{author}{\bibfnamefont{K.}~\bibnamefont{Habicht}}, \bibnamefont{and}
  \bibinfo{author}{\bibfnamefont{P.}~\bibnamefont{Vorderwisch}},
  \bibinfo{journal}{Nature} \textbf{\bibinfo{volume}{423}}, \bibinfo{pages}{62}
  (\bibinfo{year}{2003}).

\bibitem[{\citenamefont{Oosawa et~al.}(2002)\citenamefont{Oosawa, Takamasu,
  Tatani, Abe, Tsujii, Suzuki, Tanaka, Kido, and Kindo}}]{Oosawa02}
\bibinfo{author}{\bibfnamefont{A.}~\bibnamefont{Oosawa}},
  \bibinfo{author}{\bibfnamefont{T.}~\bibnamefont{Takamasu}},
  \bibinfo{author}{\bibfnamefont{K.}~\bibnamefont{Tatani}},
  \bibinfo{author}{\bibfnamefont{H.}~\bibnamefont{Abe}},
  \bibinfo{author}{\bibfnamefont{N.}~\bibnamefont{Tsujii}},
  \bibinfo{author}{\bibfnamefont{O.}~\bibnamefont{Suzuki}},
  \bibinfo{author}{\bibfnamefont{H.}~\bibnamefont{Tanaka}},
  \bibinfo{author}{\bibfnamefont{G.}~\bibnamefont{Kido}}, \bibnamefont{and}
  \bibinfo{author}{\bibfnamefont{K.}~\bibnamefont{Kindo}},
  \bibinfo{journal}{Phys. Rev. B} \textbf{\bibinfo{volume}{66}},
  \bibinfo{pages}{104405} (\bibinfo{year}{2002}).

\bibitem[{\citenamefont{Glazkov et~al.}(2004)\citenamefont{Glazkov, Smirnov,
  Tanaka, and Oosawa}}]{Glazkov04}
\bibinfo{author}{\bibfnamefont{V.~N.} \bibnamefont{Glazkov}},
  \bibinfo{author}{\bibfnamefont{A.~I.} \bibnamefont{Smirnov}},
  \bibinfo{author}{\bibfnamefont{H.}~\bibnamefont{Tanaka}}, \bibnamefont{and}
  \bibinfo{author}{\bibfnamefont{A.}~\bibnamefont{Oosawa}},
  \bibinfo{journal}{Phys. Rev. B} \textbf{\bibinfo{volume}{69}},
  \bibinfo{pages}{184410} (\bibinfo{year}{2004}).

\bibitem[{\citenamefont{Paduan-Filho et~al.}(2004)\citenamefont{Paduan-Filho,
  Gratens, and Oliveira}}]{PaduanFilho04}
\bibinfo{author}{\bibfnamefont{A.}~\bibnamefont{Paduan-Filho}},
  \bibinfo{author}{\bibfnamefont{X.}~\bibnamefont{Gratens}}, \bibnamefont{and}
  \bibinfo{author}{\bibfnamefont{N.~F.} \bibnamefont{Oliveira}},
  \bibinfo{journal}{Phys. Rev. B} \textbf{\bibinfo{volume}{69}},
  \bibinfo{pages}{020405R} (\bibinfo{year}{2004}).

\bibitem[{Bat()}]{Batista05}
\bibinfo{note}{C. D. Batista, to be published}.

\bibitem[{\citenamefont{Paduan-Filho et~al.}(1981)\citenamefont{Paduan-Filho,
  Chirico, Joung, and Carlin}}]{PaduanFilho81}
\bibinfo{author}{\bibfnamefont{A.}~\bibnamefont{Paduan-Filho}},
  \bibinfo{author}{\bibfnamefont{R.~D.} \bibnamefont{Chirico}},
  \bibinfo{author}{\bibfnamefont{K.~O.} \bibnamefont{Joung}}, \bibnamefont{and}
  \bibinfo{author}{\bibfnamefont{R.~L.} \bibnamefont{Carlin}},
  \bibinfo{journal}{J. Chem. Phys.} \textbf{\bibinfo{volume}{74}},
  \bibinfo{pages}{4103} (\bibinfo{year}{1981}).

\bibitem[{\citenamefont{Affleck}(1991)}]{Affleck91}
\bibinfo{author}{\bibfnamefont{I.}~\bibnamefont{Affleck}},
  \bibinfo{journal}{Phys. Rev. B} \textbf{\bibinfo{volume}{43}},
  \bibinfo{pages}{3215} (\bibinfo{year}{1991}).

\bibitem[{\citenamefont{Wang and Wang}(2005)}]{Wang05}
\bibinfo{author}{\bibfnamefont{H.-T.} \bibnamefont{Wang}} \bibnamefont{and}
  \bibinfo{author}{\bibfnamefont{Y.}~\bibnamefont{Wang}},
  \bibinfo{journal}{Phys. Rev. B} \textbf{\bibinfo{volume}{71}},
  \bibinfo{pages}{104429} (\bibinfo{year}{2005}).

\end{thebibliography}

\newpage

\end{document}